

\magnification 1200
\def\dslash{D \kern-6pt /}

\def\half{{1 \over 2}}
\def\p1{p_{{1 \over 2}}}

\def\l{\lambda}
\def\k{\kappa}
\def\d{\delta}
\def\m{\mu}
\def\tr{\hbox{tr}}
\def\s{\sigma}

\def\de{\delta}

\def\a{ \alpha}

\def\e{\eta}
\def\I{{\cal I}}
\def\p3{p_{{3 \over 2}}}
\def \lwd#1{\lower2pt\hbox{$\scriptstyle #1$}}

\def\h{\hbar}
\def\G{{\cal G}}
\overfullrule=0pt

\line{\hfill DIAS-STP 95-32}

\line{\hfill hep-th/9509125}

\line{\hfill September 1995}

\line{\hfill}

\centerline {{\bf Canonical Quantization}}
\centerline {{\bf of Interacting WZW Theories}}

\line{\hfill}

\centerline{C.Ford and L.O'Raifeartaigh}

\line{\hfill}

\centerline{\sl Dublin Institute for Advanced Studies}
\centerline{\sl 10 Burlington Road, Dublin 4, Ireland.}

\line{\hfill}
\centerline {\bf Abstract}

\line{\hfill}

Using canonical quantization we find the Virasoro centre for a
class of conformally-invariant interacting Wess-Zumino-Witten
theories. The theories have a group structure similar to that of
Toda theories (both abelian and non-abelian) but the usual Toda
constraints on the coupling constants are relaxed and the theories
are not necessarily integrable. The general formula for the
Virasoro centre is compared to that derived by BRST methods in the
Toda case, and helps to explain the structure of the latter.
\vskip 0.3truecm\noindent {\bf 1. Introduction}\vskip 0.2truecm

In this paper we consider the canonical quantization of a  class
of conformally-invariant interacting WZW theories.
We shall call these theories Toda-like interacting WZW theories as the
prototypes are the (abelian and non-abelian) Toda theories and
the more general ones obtained are from these by relaxing the constraints
 on the
coupling constants (see below).
 While the theories so obtained
are conformally invariant  because of the exponential form of the
 potential, there is no reason to believe that they are integrable.
For example, consider the (abelian) Toda Lagrangians
$${\cal L}=\k_{ab}\partial_\mu\phi^a\partial^\mu\phi^b-
\sum_a \mu_a^2 e^{\phi^a},\eqno(1.1)$$
where the inverse of $\k_{ab}$ is proportional to $\langle \a_a,\a_b \rangle$,
and the $\a_a$ are the simple roots of a simple Lie algebra $G$.
Now if we let $\k_{ab}$ be an arbitrary (symmetric positive definite)
 matrix   we lose integrabilty, but
we still have a conformal field theory.

We wish to consider the canonical quantization of Toda-like
theories for a number of reasons. First they provide an example
of interacting systems, which, although still two-dimensional and
conformally invariant, are more general than the systems
 previously considered.
Second, because the quantization is canonical, it provides
a useful alternative to the BRST  quantization in the Toda case.
The structure of the Virasoro central charge for the
Toda-like theories explains the structure obtained previously
[1,2,3,4] in the Toda case. Finally, because
the Toda-like theories are not first-class constrained versions
of free theories, the BRST method used [3,5]  for the Toda theories
themselves is no longer applicable.

For simplicity, we shall actually consider here only Toda-like theories
for $sl(N)$ Toda theories with integral $sl(2)$ embeddings [3,6].
This includes the abelian $sl(N)$ case, which
corresponds to the principal $sl(2)$ embedding, which is integral.
However, it will be clear from the discussion that the procedure generalizes
from $sl(N)$ to any simple Lie group and it may well generalize to
half-integral embeddings, though we have not investigated this question.

To describe the $sl(N)$ Toda-like theories we first recall that
the integral $sl(N)$ Toda theories are defined by Lagrangians of the
form
$${\cal L}=\k{\cal L}^W(g_o)-2\hbox{tr}\bigl(g_oM_+g_o^{-1}M_-\bigr) =
\k\left[\sum_\alpha{\cal L}^W(g_\alpha)+{\cal L}^W(g_{c})\right]
-2\hbox{tr}\bigl(g_oM_+g_o^{-1}M_-\bigr), \eqno(1.2)$$
where $\{M_o,M_{\pm}\}$ are the standard generators of
an embedded $sl(2)$, the (reducible) algebra
$G_o=G_c \oplus \sum_\alpha \oplus G_\alpha$ is the little algebra of $M_o$
in $sl(N)$,
with centre $G_c$, and the ${\cal L}^W$ are free
WZW Lagrangians (see section 2)
for the subalgebras indicated. The simple subalgebras
$G_\alpha$ are all $sl(n_\alpha)$ subalgebras, where $\sum_\alpha
n_\alpha=N$.

Against this background the $sl(N)$ Toda-like theories are described
by the Lagrangians
$${\cal L}=\sum_\alpha \k_\alpha{\cal L}^W(g_\alpha)+
\sum_{ab} \k_{ab}\partial
\phi^{a} \partial \phi^{b}
-2\hbox{tr}\bigl(g_oN_+g_o^{-1}N_-\bigr),  \eqno(1.3)$$
where the $\k_\alpha$ and $\k_{ab}$ are arbitrary coupling
constants, the $N_{\pm}$ are any constant matrices such that
$[M_o,N_{\pm}]=\pm N_{\pm}$.
This means that the $N_{\pm}$ intertwine only neighbouring $g_\alpha$'s
and  thus can be expanded in the form
$N_{\pm}=\sum_{a}\oplus N^{(a)}_{\pm}$
where $N_{\pm}^{(a)}$ intertwines  $g_\alpha$ and $g_{\alpha+1}^{-1}
$  for $\a=a$. Throughout we shall assume the correlation
$a\sim(\a,\a+1)$.
The weak non-degeneracy condition
$N^{(a)}_{\pm}\not=0$ for any $a$, is also assumed.

We see therefore that the  Toda-like theories are
obtained from the Toda theories by keeping the little subalgebra
$G_o$ fixed but relaxing the conditions on the coupling constants
i.e. not requiring
that the coefficients $\k$ of the kinetic terms be uniform or that the
$N_{\pm}$  be generators of the embedded $sl(2)$. We shall see that the
relaxation of the coupling constants  is the important one.
Indeed the relaxation $M_{\pm}\rightarrow N_{\pm}$ plays a role only
in so far as the $N_{\pm}$ are required to be non-degenerate.

What we mean by quantization of these theories is showing that, in
the Hilbert space of the quantum Kac-Moody current $J(x)$ of the
free WZW theory, a potential $V\bigl(g(x)\bigr)$ can be defined
in a meaningful way, and there exists a Virasoro operator of the
usual form
$$L(x)=T(x)+V\bigl(g(x)\bigr)+I(x),  \eqno(1.4)$$
where $T(x)$ is the Virasoro for the free WZW theory with
(reducible) algebra $G_o$ and $I(x)$ is an
improvement term of the form $\hbox{tr}\bigl(\I \,J'(x)\bigr)$,
 where $\I$ is a
constant matrix. As in previous models, the improvement term
$I(x)$ is introduced because the closure
of the Virasoro algebra requires that the
potential $V\bigl(g(x)\bigr)$  have conformal weights unity. However, with
respect
to $T(x)+V(x)$, it has zero weight classically and the
components $V_a$ in $V=\sum V_a$ acquire anomalous conformal
weights $\h \lambda_a$ in the quantum theory. The improvement term
is chosen so that $V\bigl(g(x)\bigr)$ has conformal weight one with
respect to $L(x)$.  We show that for the
Toda-like theories a suitable improvement term always exists, and
that the matrix $\I$ takes the form $\I=M_o-\hbar M_q$,
where $M_o$ and $\hbar M_q$ compensate  for the
classical and anomalous discrepancies $\h\lambda_a$ in the conformal weight,
respectively.  We shall compute both $\lambda_a$ and $M_q$ explicitly.

We shall show that for Toda-like potentials, the Virasoro centre
takes the form
$$C=C_{free} +24\pi\sum_{ab}
\k_{ab}\hbox{tr}\bigl(\sigma^{a}\I\bigr)
\hbox{tr}\bigl(\sigma^{b}\I\bigr),\eqno(1.5)$$
 where $C_{free}$ is the pure WZW contribution to the central charge.
We shall also show how, in the Toda case,  the formula (1.5) reduces to the
one obtained previously [3]   using BRST methods.

{}From the technical point of view the difficulty in quantizing the
Toda-like theories is that, in the non-abelian case, the
usual method of Fourier transforming to momentum-space in
order to define the Fock space via normal-ordering becomes
very complicated. What we shall show, however, is that a
decomposition into positive and negative frequencies is sufficient
to define the Fock space and to obtain our results.

\line{\hfill}

\line{\bf 2. Review of the WZW theory\hfill}

\line{\hfill}

Consider the WZW action [7] (for a  comprehensive review see [8])
$$\k\int d^2x {\cal L}^W(g)={\k}\left[\int d^2x\,\tr(g^{-1}\partial_\m
g)^2+{2\over3}\epsilon^{\alpha\beta\gamma}\int d^3x\,\tr
(g^{-1}\partial_\alpha g)(g^{-1}\partial_\beta
g)(g^{-1}\partial_\gamma g)\right],
\eqno(2.1)$$
The chiral conserved currents are defined by
$$J=\k (\partial_+ g)g^{-1},
\quad \hbox{and}\quad \bar J=-\k g^{-1}\partial_-
g,\eqno(2.2)$$
where $x^\pm=\hbox{${1\over2}$}(t\pm x)$, and
$\partial_\pm=\partial/\partial x^\pm=\partial_t\pm\partial_x$.
These currents lie in a Lie algebra $G$.
The \sl equal time \rm Poisson brackets generate a pair of commuting Kac-Moody
algebras
$$\eqalign{\{J_a(x),J_b(y)\}=&{f_{ab}}^cJ_c(x)\de(x-y)+{\k}
\e_{ab}\de'(x-y),
\cr
\{\bar J_a(x),\bar J_b(y)\}=&{f_{ab}}^c\bar J_c(x)
\de(x-y)-{\k}\e_{ab}\de'(x-y),
\cr
\{J_a(x),\bar J_b(y)\}=&0,\cr}               \eqno(2.3)
$$
where $J(x)=J_a(x)\s^a$, and the $\s^a$ matrices are understood to be in
the \sl defining \rm representation of the Lie algebra $G$, and satisfy
the  commutation relations
$$[\s^a,\s^b]=2{f^{ab}}_c\s^c.\eqno(2.4)$$
$\eta^{ab}$ is the Cartan metric
$$ \e^{ab}=\langle \s^a,\s^b\rangle=\hbox{tr}\,\s^a\s^b,\qquad
\e^{ab}\e_{bc}={\de^a}_c,\eqno(2.5)$$
which is used to raise and lower indices in the usual way.
Classically we can define  Virasoro functions
$$T(x)={\langle J(x),J(x)\rangle\over{2\k}}=
{\e^{ab}J_a(x)J_b(x)\over{2\k}},
\quad \bar T(x)=
{\e^{ab}\bar J_a(x)\bar J_b(x)\over{2\k}},\eqno(2.6)$$
which satisfy commuting Virasoro algebras
$$\eqalign{\{T(x),T(y)\}=&2T(x)\de'(x-y)+T'(x)\de(x-y),\cr
           \{\bar T(x),\bar T(y)\}=&-2\bar T(x)\de'(x-y)
-\bar T'(x)\de(x-y),                                           \cr
\{T(x),\bar T(y)\}=&0,\cr}
\eqno(2.7)$$
where the prime $'$ refers to the spatial derivative $\partial_x$.

In the quantum theory, eqs. (2.3) are replaced by the fundamental
(equal time)
commutation relations
$$\eqalign{[J_a(x),J_b(y)]=&
i\h{f_{ab}}^cJ_c(x)\de(x-y)+i\h\k \e_{ab}\de'(x-y),\cr
[\bar J_a(x),\bar J_b(y)]=&
i\h{f_{ab}}^cJ_c(x)\de(x-y)-i\h\k \e_{ab}\de'(x-y),\cr
[J_a(x),\bar J_b(y)]=&0.\cr}\eqno(2.8)$$
At the quantum level we must use normal ordered operators (or else
perform some other regularization).
Now consider the operator (which is proportional to the Virasoro
operator $T(x)$ for a pure WZW theory)
$$
t(x)=\e^{ab}:J_a(x)J_b(x):.
\eqno(2.9)
$$
The basic commutators of normal ordered products are derived in
appendix A; here we just quote the results.
Now
$$[t(x),J_p(y)]=
i\h\left[2\k J_p(x)+{\h\over{2\pi}}\e^{ab}{f_{ap}}^c{f_{cb}}^d
J_d(x)\right]\de'(x-y).\eqno(2.10)$$
 If the group is \sl simple \rm  or \sl abelian \rm
 eqn. (2.10) simplifies to
$$[t(x),J_p(y)]=i\h(2\k+\h \G)J_p(x)\de'(x-y),\eqno(2.11)$$
where $\G$ defined by
$$\e^{ab}{f_{ap}}^c{f_{cb}}^d=2\pi \G{\delta_p}^d,\eqno(2.12)$$
or $\bigl[\s^a,[\s_a,p]\bigr]=-8\pi\G p$ for $p\in G$, ie.
$-8\pi\G$ is  the quadratic Casimir in the adjoint
representation. In particular, for $sl(n)$, $-8\pi\G=\half n$.
Define the Virasoro operator
$$T(x)={t(x)\over{2\k+\h \G}},\eqno(2.13)$$
then
$$[T(x),J_p(y)]=i\h J_p(x)\de'(x-y).\eqno(2.14)$$
This is just the statement that the KM current has conformal spin
one. It is straightforward to show that $T(x)$ satisfies
 the following Virasoro algebra (again see  appendix A)
$$[T(x),T(y)]=2i\h T(x)\de'(x-y)+i\h T'(x)\de(x-y)-{i\h
C\over{24\pi}}
\de'''(x-y),\eqno(2.15)$$
with the central charge
$$C=   {2\h \k\hbox{dim}
\, G\over{2\k+\h \G}}.\eqno(2.16)$$

$T(x)$ as defined by eqn. (2.13) is not the most general Virasoro
operator, consider
$$L(x)=T(x)+I(x),\eqno(2.17)$$
where the conformal \lq\lq improvement'' term $I(x)$  is defined by
$$I(x)=\langle \I,J'(x)\rangle,\eqno(2.18)$$
and ${\cal I}\in G$. Now $L(x)$ still satisfies the Virasoro algebra (eqn.
(2.15)), but with a modified central charge
$$   C={2\h \k \hbox{dim } G\over{2\k+\h \G}}+24\pi\k
\langle \I,\I\rangle.\eqno (2.19)$$

\line{\hfill}
\line{\bf 3. Conformal character of g(x)\hfill}

\line{\hfill}
When considering interactions, we will need the Poisson bracket (and
quantum commutator) relations regarding $g(x)$.
Classically $g(x)$ satisfies
$$2\k g'(x)=(\partial_+-\partial_-)g=J(x)g(x)+g(x)\bar J(x).
\eqno(3.1)$$
Together with appropriate boundary conditions, we can regard eqn.
(3.1) as defining $g(x)$ in terms of the KM variables.
$g(x)$ obeys the following Poisson bracket relations
$$\{J_p(x),g(y)\}=-\hbox{$\half$}\s_pg(y)\d(x-y),\qquad
 \{\bar J_p(x),g(y)\}=\hbox{$\half$}g(y)\s_p\d(x-y),
\eqno(3.2)$$
and
$$\{[g(x)]_{ab},[g(y)]_{cd}\}=0.\eqno(3.3)$$
Equation (3.3) is sometimes written as
$$\{g(x)\,\,,^{\!\!\!\otimes}g(y)\}=0.\eqno(3.4)$$
Note that the $\{g \,\,,^{\!\!\!\otimes}g\}$ bracket is non-trivial
on the light cone, and involves the Yang-Baxter $r$-matrices
[9].

We cannot transplant eqn. (3.1) directly into the quantum
theory since it is not a normal ordered equation.
The quantum analogue of eqn. (3.1) is [10]
$$\beta g'(x)=\s^a:J_a(x)g(x):+:\bar J_a(x)g(x):\s^a,\eqno(3.5)$$
where
$$\beta=2\k+\h \G.\eqno(3.6)$$
We show in  appendix B that eqn. (3.5) together with the KM algebra eqn.
(2.8) is compatible with the basic commutators
$$[J_p(x),g(y)]=-\hbox{${i\over2}$}\h\s_p g(y)\d(x-y),\qquad
[\bar J_p(x),g(y)]=\hbox{${i\over2}$}\h g(y)\s_p \d(x-y),\eqno(3.7)$$
and
$$[g(x)\,\,,^{\!\!\!\otimes}g(y)]=0.\eqno(3.8)$$
If one attempted to define a $g(x)$ operator with a value of $\beta$
other than that given by eqn. (3.6) the $[J_p(x),g(y)]$ commutator
would be non-local in the non-abelian case.

We can also define a normal ordered $g^{-1}(x)$ operator via
$${\beta}\partial_x g^{-1}(x)=
-:J_a(x)g^{-1}(x):\s^a-\s^a:\bar J_a(x)g^{-1}(x):,\eqno(3.9)$$
and this satisfies
$$[J_p(x),g^{-1}(y)]=\hbox{${i\over2}$} \h g^{-1}(y)\s_p
\delta(x-y),\qquad
        [\bar J_p(x),g^{-1}(y)]=-\hbox{${i\over2}$}\h
        \s_p  g^{-1}(y)\delta(x-y).
      \eqno(3.10)$$
Note that $g^{-1}(x)$ is \sl not \rm the inverse of $g(x)$ (except
in the limit $\hbar\rightarrow 0$), since the inverse of $g(x)$
would be anti-normal ordered.

Classically we have
$$\{T(x),g(y)\}+\{g(x),T(y)\}=0,\eqno(3.11)$$
which is just the statement that $g(x)$ is a conformal scalar.
In the quantum theory $g(x)$ acquires an anomalous dimension
(see Appendix B); the
analogue of eqn. (3.11) is
$$[T(x),g(y)]+[g(x),T(y)]=-{i\h^2c\over{8\pi(2\k+\h \G)}}\Bigl[2g(x)
\de'(x-y)+g'(x)\de(x-y)\Bigr],\eqno(3.12)$$
where   $c$ is the quadratic Casimir in the defining representation,
ie.
$$c{I}=\eta^{ab}\s_a\s_b,\eqno(3.13)$$
$I$ being the identity matrix,
and $T(x)$ is the correctly normalized kinetic term given by
eqn. (2.13).  We also have
$$[T(x),g^{-1}(y)]+[g^{-1}(x),T(y)]=-{i\h^2c\over{8\pi(2\k+\h \G)}}
\Bigl[2g^{-1}(x)
\de'(x-y)+\partial_x{g^{-1}}(x)\de(x-y)\Bigr].\eqno(3.14)$$
Note that the sign on the RHS of eqn. (3.14) is the \sl same \rm sign as
on the RHS of eqn. (3.12), whereas if $g^{-1}(x)$ was the inverse of
$g(x)$ (rather than just a normal ordered version of the inverse)
the signs would be opposite.

\vfill\eject

\line{\bf 4. Classical Interaction Potentials\hfill}
\line{\hfill}

 We now  consider \sl interacting \rm
WZW models which are conformally invariant.
  Let us add a \lq\lq
potential'' term to the ordinary WZW Lagrangian eqn. (2.1).
$${\cal L}={\cal L}^W(g)-2 V(g).\eqno(4.1)$$
Classically, this will entail
adding a potential term to the canonical energy momentum tensor
$$T^{\m\nu}_{can}(x)=T^{\m\nu}_{W}(x) +2g^{\m\nu}V(x),
\eqno(4.2)$$
where  $T^{\m\nu}_{W}$ refers to the free WZW canonical energy momentum
tensor.
One can see immediately that \sl any \rm potential term will spoil
the tracelessness of $T^{\m\nu}_{can}(x)$.
Moreover, it is not possible to construct a Virasoro algebra
based on the canonical EM tensor, ie. $L_{can}(x)=
\half[T_{can}^{00}(x)+T_{can}^{01}(x)]$ and $\bar L_{can}(x)=\half
[T_{can}^{00}(x)-T_{can}^{01}(x)]$ do not satisfy the Virasoro
algebra. To see this, consider
$$L_{can}(x)=T(x)+V(x),\qquad
\bar L_{can}(x)=\bar T(x)+V(x),\eqno(4.3)$$
where $T(x)$ and $\bar T(x)$ refer to the canonical pure WZW
contribution. We assume that
$$\{V(x),V(y)\}=0,\eqno(4.4)$$
ie. we suppose that our potential $V(x)$ contains no derivatives of
$g(x)$, and then eqn. (4.4) follows from eqn. (3.4).
Now since
$$\{T(x),V(y)\}+\{V(x),T(y)\}=0,\eqno(4.5)$$
which is equivalent to the statement that $V(x)$ has conformal spin zero,
then
$$\eqalign{\{L(x),L(y)\}=&2T(x)\de'(x-y)+T'(x)\de(x-y)\cr
\neq& 2L(x)\de'(x-y)+L'(x)\de(x-y)-{C\over{24\pi}}\de'''(x-y).
\cr}\eqno(4.6)$$
However, we know from section 2 that the canonical EM tensor is not
the most general EM tensor for the \sl pure \rm WZW theory which
generates a Virasoro algebra. In the pure WZW case we are free
to add an improvement term $I(x)$ whose only effect is to modify the
central charge $C$. As alternatives to $L_{can}$ and $\bar L_{can}$
consider the following Virasoro functions (such a procedure was
used by Curtright and Thorn to improve the Liouville
energy-momentum tensor
at the classical and quantum level [11], see also [12])
$$L(x)=T(x)+V(x)+I(x),\qquad \bar L(x)=\bar T(x)+V(x)+\bar I(x),
\eqno(4.7)$$
where as in section 2, $I(x)$ and $\bar I(x)$ are understood to be
linear in the spatial derivatives of the KM currents,
with the Poisson bracket relations
$$\{I(x),I(y)\}=-{C\over{24\pi}}\de'''(x-y),\qquad
                         \{\bar I(x),\bar
I(y)\}=
{C\over{24\pi}}\de'''(x-y),$$
$$\{T(x),I(y)\}=I(x)\de'(x-y),\qquad
\{\bar T(x),\bar I(y)\}=-\bar I(x)\de'(x-y).\eqno(4.8)$$
Now
$$\eqalign{\{L(x),L(y)\}=&2\bigl(T(x)+I(x)\bigr)\de'(x-y)+
\bigl(T'(x)+I'(x)\bigr)\de(x-y)\cr
&+\{L(x),V(y)\}+\{V(x),L(y)\}+\{I(x),I(y)\}\cr
=&2L(x)\de'(x-y)+L'(x)\de(x-y)-{C\over{24}}\de'''(x-y),\cr
}\eqno(4.9)$$
if
$$\{L(x),V(y)\}+\{V(x),L(y)\}=
2V(x)\de'(x-y)+V'(x)\de(x-y).\eqno(4.10)$$
Using eqn. (4.5) this simplifies to
$$\{I(x),V(y)\}+\{V(x),I(y)\}=
2V(x)\de'(x-y)+V'(x)\de(x-y).\eqno(4.11)$$
or
$$\{I(x),V(y)\}=V(y)\de'(x-y).\eqno(4.12)$$
That is if we can find an improvement term that satisfies the above
equation (there will be a similar condition concerning $\bar I(x)$)
then we have a conformally invariant interaction. The classical
central extension of the Virasoro algebra is entirely due to the
$\{I(x),I(y)\}$ bracket.

\line{\hfill}

\noindent{\bf 5. Quantum Interaction Potentials}

\line{\hfill}
We extend the considerations of the previous section to the quantum
level. As in the classical case,
 the task of
establishing conformal invariance and computing the central charge
amounts to the problem of finding suitable improvement operators
$I(x)$ and $\bar I(x)$. However there is a slight complication, in
that
$[T(x),V(y)]+[V(x),T(y)]$ is in general non-zero (this follows from eqn.
(3.12)). From eqn. (3.8) (and the assumption that our interaction
potential contains no derivatives) we have
$$[V(x),V(y)]=0,\eqno(5.1)$$
and so the conformal invariance condition analogous to eqn. (4.10)
is
$$[L(x),V(y)]+[V(x),L(y)]=2i\h V(x)\de'(x-y)+i\h V'(x)\de(x-y).
\eqno(5.2)$$
Let us assume, however, that the potential can be \lq\lq diagonalized''
  with
respect to the action of $T(x)$, ie.
$$V(y)=\sum_a V_a(y),\eqno(5.3)$$
where each of the components $V_a$ have a definite conformal weight
$\h\lambda_a$,
$$[T(x),V_a(y)]+[V_a(x),T(y)]=i\h^2\lambda_a\left[
2V_a(x)\de'(x-y)+V_a'(x)\de(x-y)\right].\eqno(5.4)$$
There is no summation over $a$ in the last equation.
We will demonstrate such a decomposition for Toda-like theories in
the next section.
Substituting eqn. (5.4) into eqn. (5.2), the conformal invariance
condition reduces to
$$[I(x),V_a(y)]+[V_a(x),I(y)]= i\h(1-\h\l_a)\left[
  2  V_a(x)\de'(x-y)+V_a'(x)\de(x-y)\right],\eqno (5.5)$$
    or
$$[I(x),V_a(y)]=i\h(1-\h\l_a)V_a(y)\de'(x-y),\eqno(5.6)$$
which is the quantum analogue of eqn. (4.12).
 Thus, the problem reduces to finding an improvement term $I(x)$
which satisfies eqn. (5.6). In the next section we will
show that for \lq\lq Toda-like'' theories this problem is easily
solved.

\vskip 0.3truecm \noindent {\bf 6. Toda-like Theories}
\vskip 0.2truecm

We now consider the Toda-like Lagrangians defined by eqn. (1.3).
As mentioned in the introduction, the $G_o$ algebra splits into a
central piece $G_c$ and a non-abelian piece $\sum_\alpha G_\alpha$
which is just the sum of the $sl(n_\alpha)$ components. It is
convenient to use a (non-orthogonal) basis for $G_c$ with the
properties
$$[\sigma_a,M_\pm]=\pm M_\pm^{(a)},\quad  \hbox{and}\quad
M_o=\sum_a\s_a.\eqno(6.1)$$
The explicit form of the $\sigma_a$ matrices and their duals
$\sigma^a$ is given in appendix C.
Note that the $\s_a$ matrices also satisfy
$$[\sigma_a,N_\pm]=\pm N_\pm^{(a)}.\eqno(6.2)$$
We can write, $g_c(x)$, the central piece of $g_o(x)$ as follows
$$g_c(x)=e^{\phi^a(x)\s_a}.\eqno(6.3)$$
The potential term in eqn. (1.3) can be decomposed as follows
$$V\bigl(g_o(x)\bigr)=\hbox{tr}\left(g_oN_+g_o^{-1}N_-\right)=
\sum_aV_a(g_o),\eqno(6.4)$$
where
$$V_a(g_o)=\hbox{tr}\left(g_\alpha N_+^{(a)}g^{-1}_{\alpha+1}N_-^{(a)}
\right)
e^{\phi^a},\quad(\a=a),\eqno(6.5)$$
where $\alpha=a$, and the $g_\alpha$ are the $sl(n_\alpha)$ WZW variables.
 The central fields $\phi^a(x)$
 satisfy the Poisson bracket relation
$$\{j_b(x),\phi^a(y)\}=-\hbox{$\half$}{\delta_b}^a\delta(x-y),
\eqno(6.6)$$
which follows from eqn. (3.2) applied to $G_c$. The central KM
currents satisfy
$$\{j_a(x),j_b(y)\}=\k_{ab}\delta'(x-y).\eqno(6.7)$$

When normal ordering the $V_a$ potentials, one can normal order
the $g_\alpha$, $g_{\alpha+1}^{-1}$ and $e^{\phi^a}$ separately using
the methods described in appendix B.
 The free quantum Virasoro operator takes
the usual form
$$T(x)=\sum_\alpha {1 \over
2\k_\alpha+\h\G_\alpha}\hbox{tr}\left[:J\bigl(g_\alpha(x)\bigr)^2:\right]
+{1 \over 2}\sum_{ab}\k^{ab}:j_{a}(x)j_{b}(x):\eqno(6.8)$$
where
 $\k^{ab}$ is the inverse of $\k_{ab}$.
The decomposition $V=\sum_aV_a$ given in eqn. (6.4) is already of the
\lq\lq diagonal'' form, in that the $V_a$ satisfy eqn. (5.4), with
$$8\pi\lambda_a=-{1\over2} \k^{aa}
-{ c_\alpha \over 2\k_\alpha+\hbar \G_\alpha}
-{ c_{\alpha+1} \over 2\k_{\alpha+1}+\hbar \G_{\alpha
+1}},\quad(\a=a).  \eqno(6.9)$$
Here, the $-c_\a/(2\k+\h\G_\a)$ term is due to the anomalous conformal
weight of $g_\alpha(x)$ (see eqn. (3.12)), the $-c_{\a+1}/
(2\k+\h\G_{\a+1})$ is the contribution from $g_{\alpha+1}^{-1}$ (eqn.
(3.14)),
and the $-\half\k^{aa}$ term is the due to the (abelian)
$:e^{\phi^\a}:$ term in the normal ordered version of eqn. (6.5).
For $sl(n_\alpha)$ we have $\G_\a=-n_\alpha/(16\pi)$, and
\footnote{$\dagger$}{Let $\s_a$ be a basis for $sl(n_\alpha)$ in
the defining representation. Taking the trace of $c_\alpha{I}=\s^a\s_a$
gives $n_\alpha c_\alpha=\hbox{dim}\,G_\alpha=n_\alpha^2-1$.}
$c_\alpha=(n_\a^2-1)/n_\alpha$.
For principal embeddings (abelian Toda-like theories) we have
$c_\alpha=c_{a+1}=0$ and thus $\lambda_a$ reduces to
$-\k^{aa}/(16\pi)$. Note that even in this case the $\lambda_a$ are
not, in general, uniform in $a$ and thus the components $V_a$ of the
potential have different anomalous conformal weights.

The system is conformally invariant if there exists an $I(x)$ which
satisfies eqn. (5.6).
It is easy to see that  $I(x)=\langle \I,J'(x)\rangle$, with
$$\I=\sum_{a}(1-\h\lambda_a)\sigma_{a}=M_o-\hbar M_q,
\eqno(6.10)$$
is a solution to eqn. (5.6).
The centre of the Virasoro algebra  is evidently
$$C=\hbar\left[\hbox{dim}\,G_c+\sum_\alpha {2\k_\alpha \hbox{dim}\,
G_\alpha \over
2\k_\alpha+\h\G_\alpha}\right]
+24\pi\sum_{ab}\k_{ab}\hbox{tr}\bigl(\sigma^{a}I\bigr)
\hbox{tr}\bigl(\sigma^{b}I\bigr)    \eqno(6.11)$$
In the abelian case this reduces to
$$C=\hbar\,\hbox{dim}\,H
+24\pi\sum_{ab}\k_{ab}\hbox{tr}\bigl(\sigma^{a}\I\bigr)
\hbox{tr}\bigl(\sigma^{b}\I\bigr),   \eqno(6.12)$$
$H$ being the Cartan subalgebra of $G$.
\vskip 0.3truecm \noindent {\bf 7. Comparison with the Reduction Formula}
\vskip 0.2truecm

Classically, the Toda case is obtained from the general one by
imposing the uniformity relations $\k_{ab}=\k\,\tr(\s_a\s_b)$
and $\k_\alpha=\k$. At the quantum level, however, the uniformity
conditions are [3,5]
$$2\k_{ab}=(2\k+\hbar \G)\hbox{tr}(\sigma_{a}\sigma_{b})\quad
\hbox{and} \quad 2\k_\alpha+\hbar \G_\alpha=2\k+\hbar \G,  \eqno(7.1)$$
 where $-8\pi\G$ is the quadratic Casimir for $sl(N)$ in the adjoint
representation, ie. $\G=-N/(16\pi)$.

 On using the conditions (7.1) the expression (6.10) for
the central charge simplifies to
$$C=\hbar\left[\hbox{dim}G_c+\sum_\alpha {2\k_\alpha\hbox{dim}G_\alpha \over
2\k_\alpha+\h\G_\alpha}\right]
+12\pi(2\k+\hbar \G)\hbox{tr}\Bigl[\bigl(M_o-\hbar
M_q\bigr)^2\Bigr],
  \eqno(7.2)$$
where
$M_o-\hbar M_q = \sum_a\left(1-\h\lambda_a\right)\sigma_a
  $.
Using the expression for $\s^a$ given in appendix C we have
$$\k^{aa}={2 \over (2\k+\hbar \G)}\hbox{tr}(\sigma^a\sigma^a)
={2 \over (2\k+\hbar \G)}\left({1 \over n_{\alpha}}+{1 \over
n_{\alpha+1}}\right), \quad(\alpha=a), \eqno(7.3)$$
and
$c_\alpha=(n_\alpha^2-1)/n_\alpha$.
Therefore eqn. (6.9) for $\lambda_a$ simplifies to
$$8\pi\lambda_a=-{1\over{ (2\k+\hbar \G)}}
\bigl(n_{\alpha}+n_{\alpha+1}\bigr),\quad(\alpha=a).  \eqno(7.4)$$

 The formula obtained previously using BRST methods is [3]
$$C=\hbar \hbox{dim}G_o+12\pi(2\k+\hbar \G)\hbox{tr}\left[
M_o+{\hbar m_o \over
4\pi(2\k+\hbar \G)}\right]^2,  \eqno(7.5)$$
where $M_o$ and $m_o$ are the grading operators corresponding to
the actual and the principle $sl(2)$ embeddings respectively. The
formulae (7.2) and (7.5) are rather similar but they cannot be
compared immediately because (7.5) is not in the form in which the part
coming from the free theory is separated from the part
due to the potential. However, (7.5) can be converted to the
canonical form by decomposing  $m_o$ into its
block components. Let $m_o^{(\alpha)}$ be the grading operators for
the principal $sl(2)$-embeddings within blocks and $\bar m_\alpha$
denote the block-average of $m_o$ within the $sl(n_\alpha)$ block.
Then, since the spectra of both $m_o$ and $m_o^{(a)}$
are integer spaced and  the block-trace of $m_o^{(a)}$ is zero, we
have
$$m_o=\sum_\alpha m_o^{(a)}+\bar M \quad \hbox{where} \quad \bar M
=\bar m_\alpha I_\alpha.   \eqno(7.6)$$
The tracelessness of the $m_o^{(a)}$ within blocks ensures that
the $m_o^{(a)}$ are trace-orthogonal to $M_o$  and
to $\bar M$ and, using this trace-orthogonality, we can write eqn.
(7.5) in the form
$$C=\hbar\hbox{dim}G_c+ \hbar \sum _\alpha
\left[\hbox{dim}G_\alpha+{3\hbar\hbox{tr}\bigl(m_o^{(\alpha)}\bigr)^2\over
{4\pi( 2
\k_\alpha+\hbar
\G_\alpha)}}\right]
+12\pi(2\k+\hbar \G)\hbox{tr}\left[M_o+{\hbar \bar M\over{4\pi (2\k+\hbar
\G)}}\right]^2.  \eqno(7.7)$$
Using $m_o^{(\alpha)}=\hbox{diag}\left[\half
(n_\alpha-1),\half(n_\alpha-3),...,-\half(n_\alpha-1)\right]$ it is
straightforward to show that $\tr\bigl(m_o^{(\alpha)}\bigr)^2=
{1\over{12}}n_\alpha(n_\alpha+1)(n_\alpha-1)$. This can be rewritten
as $\tr\bigl(m_o^{(\alpha)}\bigr)^2={1\over{12}}n_\alpha \hbox{dim}
G_\alpha=-{4\over3}\pi\G_\alpha \hbox{dim} G_\alpha$. Inserting the
last equality into eqn. (7.7) gives
$$C=\hbar \left[\hbox{dim}G_c+\sum_\alpha {2\k_\alpha \hbox{dim}
G_\alpha\over
2\k_\alpha+\hbar \G_\alpha}\right]
+12\pi(2\k+\hbar \G)\hbox{tr}\left[M_o+{\hbar \bar M \over {4\pi
(2\k+\hbar \G)}}
\right]^2.
\eqno(7.8)$$
In (7.8) the part of the centre coming from the free WZW theory is
separated (in the square brackets) as required.

We can now compare (7.8) and (7.2) and we see that they are the
same provided that $2\bar M=-(2\k+\hbar \G)M_q$. But from (7.6) and
(7.4) we see that this holds if
$$2\sum_\alpha \bar m_\alpha I_\alpha=-\sum_a
(n_{\alpha}+n_{\alpha+1})\sigma_a.
\eqno(7.9)$$
But from the definition of $\sigma_a$ given in appendix C we have
$$\eqalign{
\sum_\alpha \bar m_\alpha I_\alpha
&=
\sum_\alpha \left[\bar m_\alpha (\sigma_a-\sigma_{a-1})
-{\bar m_\alpha n_{\alpha}\over N}I\right]   \cr
&=\sum_\alpha \bar m_\alpha (\sigma_a-\sigma_{a-1})
=\sum_\alpha (\bar m_\alpha-\bar m_{\alpha+1}) \sigma_a
=-{1 \over 2}\sum_\alpha (n_\alpha+n_{\alpha+1}) \sigma_a,   \cr }
\eqno(7.10)$$
where we have used the fact that
$\sum \bar m_{\alpha} n_{\alpha}=\hbox{tr}(\bar M)=0$.
Thus the general formula for the centre reduces to the BRST formula
in the Toda case, as required. We  also see that
the formula (7.8) and not (7.5) is the more natural one for the Toda theory.

\line{\hfill}

\noindent{\bf 8. Conclusions}

\line{\hfill}

In this paper we have considered the canonical quantization of
conformally invariant interacting WZW theories.
In particular, we have considered \lq\lq Toda-like'' models, which
for special values
of the couplings  reduce to the well known Toda (abelian and
non-abelian) theories. In the canonical approach, the computation of
the Virasoro central charge amounts to the problem of finding a
suitable \lq\lq improved'' energy-momentum tensor. In the Toda case,
our formula for
the central charge reduces to the expression previously obtained
in the BRST approach.

\line{\hfill}

\noindent{\bf Acknowledgements}

\line{\hfill}
We thank W.McGlinn, I.Sachs and C.Wiesendanger for useful discussions.
 L.O'R. is grateful to the University of Notre Dame for their kind
hospitality and financial support while much of this work was being
           performed.

\line{\hfill}

\line{\bf Appendix A: Normal Ordering\hfill}

\line{\hfill}

In order to define a Virasoro operator in the quantum case, we
introduce a normal ordering prescription.
 This is usually accomplished by identifying the Fourier modes of $J(x)$
 and $\bar J(x)$ as either creation or annihilation operators with
 respect to a vacuum state $|0\rangle$. Then an operator is normal
 ordered if all annihilation operators are placed to the right of
 creation operators. However, we can perform the splitting of $J(x)$
 into creation and annihilation operators while remaining in \sl
 position \rm space as follows.
Write
$$J(x)=J^+(x)+J^-(x),\quad \hbox{and}\quad \de(x)=\de^+(x)+\de^-(x),
 \eqno(A.1)$$
where
 $$J^\pm(x)=\int ^\infty_{-\infty} dy\, \de^\pm(x-y)J(y),\eqno(A.2)$$
 and
 $$
 \delta^\pm(x)= \int^{\infty}_{-\infty}
{dk\over{2\pi}}\theta(\mp k)e^{ik(x\mp i\epsilon)}=\pm{i\over{2\pi }}
{1\over{(x\mp i\epsilon)}},\eqno (A.3)$$
where $\epsilon>0$.
Trivially we also have $\delta^\pm(-x)=\delta^\mp(x)$ and
    $\delta^\pm {}'(-x)=-\delta^\mp {}'(x)$, where
$\delta^\pm{}'(x)=\partial_x\delta^\pm(x)$.
We now interpret $ J^-(x)$ and $J^+(x)$ as creation and annihilation
operators, respectively
$$       J^+(x)|0\rangle=0,\qquad        \langle 0   |  J^-(x)=0,
\eqno(A.4)$$
 and as usual an operator is normal ordered if all $J^+$ operators
 are placed to the right of $J^-$ operators. Similarly we decompose
 $\bar J(x)=\bar J^-(x)+\bar J^+(x)$. However the normal ordering of the
 $\bar J$ operators is done in the {\sl opposite} way to the $J$
operators;
  an operator is normal ordered if all $\bar J^-$ operators are
 placed to the right of $\bar J^+$ operators.
 We will make extensive use of the following identities, which
follow immediately from eqn. (A.3)
$$\eqalignno{
\left[\de^+(x)\right]^2-\left[\de^-(x)\right]^2=&
{i\over{2\pi}}\de'(x)&(A.5)\cr
\left[{\de^+}'(x)\right]^2-\left[{\de^-}'(x)\right]^2=&
{i\over{12\pi}}\de'''(x).&(A.6)\cr}$$
 Multiplying the KM algebra eqs. (2.8)  by $\de^\pm
(z-x)$ and integrating with respect to $x$ gives the very useful
relation
$$[J^\pm_a(z),J_b(y)]=i\h{f_{ab}}^cJ_c(y)\de^\pm(z-y)+{i\h
\k} \e_{ab}\de^\pm{}'(z-y),
\eqno(A.7)$$
and similarly for   $[\bar J^\pm_a(z),\bar J_b(y)]$.
Notice the current on the RHS of eqn. (A.7) is the full current,
and not just the creative or destructive piece.

Writing the $t(x)$ operator defined by eqn. (2.9) in terms of
$J^+(x)$ and $J^-(x)$
$$
t(x)=\e^{ab}:J_a(x)J_b(x):=\e^{ab}
\left[J_a^+(x)J_b^+(x)+J_a^-(x)J_b^-(x)+2J_a^-(x)J_b^+(x)
\right].
\eqno(A.8)$$
We derive eqn. (2.10) for the $[t(x),J_p(y)]$ commutator
$$\eqalign{[t(x),J_p(y)]=&2i\h \k J_p(x)\de'(x-y)\cr&+ i\h\e^{ab}{f_{ap}}^c
\Bigl\{ \bigl[J^+_b(x)J_c(y)+
J_c(y)J_b^+(x)+2J_b^-(x)J_c(y)\bigr]\d^+(x-y)\cr&
\,\,\,\,\,\,\,+\bigl[J_b^-(x)
J_c(y)+J_c(y)J_b^-(x)+2J_c(y)J_b^+(x)\bigr]\d^-(x-y)\Bigr\}.\cr}
\eqno(A.9)$$
Some of the terms on the right hand side of eqn. (A.9) are not
correctly normal ordered. We  rewrite them in terms
of normal ordered objects,
$$\eqalign{J^+_b(x)J_c(y)+
J_c(y)J_b^+(x)+2J_b^-(x)J_c(y)=&:J_b(x)J_c(y)+J_c(y)J_b(x):+
[J_b(x),J_c^-(y)],\cr
 J_b^-(x)
J_c(y)+J_c(y)J_b^-(x)+2J_c(y)J_b^+(x)=&
:J_b(x)J_c(y)+J_c(y)J_b(x):+[J_c^+(y),J_b(x)],\cr} \eqno(A.10)$$
and here
$$:J_b(x)J_c(y):=J_b^+(x)J_c^+(y)+J_b^-(x)J_c^-(y)+J_b^-(x)J_c^+(y)+
J_c^-(y)J_b^+(x).\eqno (A.11)$$
Thus
$$\eqalign{[t(x),J_p(y)]=
&2i\h \k J_p(x)\de'(x-y)\cr&+ i\h \e^{ab}{f_{ap}}^c\Bigl
\{\bigl(:J_b(x)J_c(y)+J_c(y)J_b(x):+[J_b(x),J_c^-(y)]\bigr)
\de^+(x-y)\cr&
\,\,\,\,\,\,\,+\bigl(:J_b(x)J_c(y)+J_c(y)J_b(x):
+[J_c^+(y),J_b(x)]\bigr)\de^-(x-y)\Bigr\}\cr=&2i\h
\k J_p(x)\d'(x-y)+
i\h\e^{ab}{f_{ap}}^c\left(:J_b(x)J_c(y)+J_c(y)J_b(x):\right)
\de(x-y)\cr&+\h^2\e^{ab}{f_{ap}}^c{f_{cb}}^d J_d(x)
\bigl\{[\d^+(x-y)]^2-[\de^-(x-y)]^2\bigr\}\cr
=&i\h\left[2\k J_p(x)+{\h\over{2\pi}}\e^{ab}{f_{ap}}^c{f_{cb}}^d
J_d(x)\right]\de'(x-y),\cr}\eqno(A.12)$$
which is eqn. (2.10). In equation (A.12) we have used the identity
(A.5).
 If the group is \sl simple \rm  or \sl abelian \rm
then eqn. (A.12) simplifies to
$$[t(x),J_p(y)]=i\h(2\k+\h \G)
J_p(x)\de'(x-y),\eqno(A.13)$$
where $\G$ is defined by eqn. (2.12).

We now compute  $[t(x),t(y)]$.
 Using the $[t(x),J(y)]$ result
it is easy to see that
$$[t(x),J_p^\pm(y)]=i\h (2\k+\h \G)J_p(x)\de^\mp{}'(x-y).\eqno(A.14)$$
Now
$$\eqalign{[t(x),t(y)]=&\e^{ab}\Bigl\{J_a^+(y)[t(x),J_b^+(y)]+
[t(x),J_a^-(y)]J_b^-
(y)
+[t(x),J_a^+(y)]J_b^+(y)\cr
&\,\,\,+ J_a^-(y)
[t(x),J_b^-(y)]+2J_a^-(y)[t(x),J_b^+(y)]+
2[t(x),J_a^-(y)]J_b^+(y)\Bigr\}.\cr}\eqno(A.15)$$
The first two entries on the RHS of the last equality are not
normal ordered, and so they must be \lq\lq re-normal ordered'' as in
eqn. (A.10).
Which gives
$$\eqalign{[t(x),t(y)]
           =&\e^{ab}\Bigl\{ 2[t(x),J_a^+(y)]J_b^+(y)+2J_a^-(y)[t(x),J_b^-(y)]+
2J_a^-(y)[t(x),J_b^+(y)]\cr &\,\,+2[t(x),J_a^-(y)]J_b^+(y)
           +\bigl[J_a^+(y),[t(x),J_b^+(y)]\bigr]-\bigl[J_a^-(y),
[t(x),J_b^-(y)]\bigr]\Bigr\}
           \cr
           =&i\h (2\k+\h \G)
\e^{ab}\left(:J_a(x)J_b(y)+J_a(y)J_b(x):\right)\de'(x-y)      \cr
                &-\h^2 \k\,(2\k+\h\G)
\hbox{dim}\, G
 \left\{[\de^+{}'(x-y)]^2-[\de^-{}'(x-y)]^2\right\}      \cr
         =&2i\h(2\k+\h \G)
t(x)\d'(x-y)+i\h(2\k+\h \G)
t'(x)\de(x-y)          \cr
             &-{i\h^2 \k\over{12\pi}}(2\k+\h \G) \,
\hbox{dim}\, G\,\de'''(x-y),
        \cr }\eqno(A.16) $$
using the identity eqn. (A.6). Using eqn. (2.13), the
Virasoro algebra eqn. (2.15) follows immediately.
We note that the central extension in eqn. (2.15) is just the contribution from
the double commutator terms that arise from \lq\lq re-normal''
ordering, ie.
$$i\h C\de'''(x-y)=24\pi \e^{ab}\Bigl\{\bigl[J_a^+(y),
[J_b^+(y),T(x)]\bigr]-\bigl[J_a^-(y),[J_b^-(y),T(x)]\bigr]\Bigr\}.
\eqno(A.17)$$

\line{\hfill}

\line{\bf Appendix B: Properties of g(x)\hfill}

\line{\hfill}

 We  now show that $g(x)$ satisfying the normal ordered equation
(3.5) is compatible with eqs. (3.7) only when $\beta=
2\k+\h \G$. In terms of $J^\pm(x)$, eqn. (3.5) reads
$$\beta g'(x)=\left[\s^aJ_a^-(x)g(x)+\s^a g(x)J_a^+(x)
+\bar J_a^+(x)g(x)\s^a+g(x)\bar J^-(x)\s^a\right].\eqno(B.1)$$
Now
$$\eqalign{\beta[J_p(z), g'(x)]=&
\s^a[J_p(z),J_a^-(x)] g(x) +
\s^aJ_a^-(x)[J_p(z), g(x)]
+\s^a[J_p(z), g(x)]J_a^+(x) \cr
&+\s^a g(x)[J_p(z),J_a^+(x)]+
\bar J_a^+(x)[J_p(z),g(x)]\s^a+[J_p(z),g(x)]\bar J_a^-(x)\s^a\cr
=&i\h \k\s_p g(x)\d'(z-x)\cr
&+i\h{f_{pa}}^d\s^aJ_d(z) g(x)\d^+(z-x)+i\h{f_{pa}}^d\s^a g(x)
J_d(z)\d^-(z-x)\cr
&+ \s^aJ_a^-(x)[J_p(z), g(x)]
+\s^a[J_p(z), g(x)]J_a^+(x)\cr
   &+\bar J_a^+(x)[J_p(z),g(x)]\s^a +[J_p(z),g(x)]\bar J_a^-(x)\s^a.
\cr
}\eqno(B.2)$$
 Re-normal ordering the  right hand side of eqn.
 (B.2),
we have
$$\eqalign{\beta\partial_x[J_p(z),g(x)]=&
i\h \k \s_p\d'(z-x)g(x)+i\h{f_{pa}}^d\s^a \left[J_d^-(z)g(x)+g(x)J_d^+(z)
\right]\d(x-z)\cr &+
i\h\s^a{f_{pa}}^d\left\{[J_d^+(z),g(x)]\d^+(z-x)+ [g(x),J_d^-(z)]
\d^-(z-x)\right\}\cr
&+\s^aJ_a^-(x)[J_p(z), g(x)]
+\s^a[J_p(z), g(x)]J_a^+(x)\cr
   &+\bar J_a^+(x)[J_p(z),g(x)]\s^a +[J_p(z),g(x)]\bar J_a^-(x)\s^a.\cr}
\eqno(B.3)$$
One can derive a similar equation for $\partial_x[\bar J_p(z),g(x)]$.
We can regard eqn. (B.3)  as a differential equation for
$[J_p(z), g(x)]$. Now we will show that  eqs. (3.7), ie.
$$[J_p(z),g(x)]=-\hbox{${i\over2}$}\h\s_p g(x)\d(z-x),\qquad
[\bar J_p(z),g(x)]=\, \hbox{${i\over2}$}\h g(x)\s_p
\d(z-x),\eqno(B.4)$$
 satisfy eqn. (B.4) for a \sl unique \rm value of $\beta$.
Inserting eqs. (B.4) into eqn. (B.3) gives
$$\eqalign{-\hbox{${i\over2}$}\h\beta\s_p\partial_x\bigl[g(x)\de(z-x)\bigr]=&
\hbox{${i\over2}$}\h(2\k+\h \G)\s_p g(x)\d'(z-x)\cr &+
                    i\h  \s^a {f_{pa}}^d
\left[J_d^-(z) g(x)+g(x)J_d^+(z)\right]\d(x-z)\cr &-
\hbox{${i\over2}$} \h\s^a\s_p
\left[J_a^-(z)g(x)+g(x)J_a^+(z)\right]\d(z-x)\cr
&-\hbox{${i\over2}$}\h\s_p\left[g(x)\bar J^-_a(x)+\bar
J^+_a(x)g(x)\right]\s^a\de(z-x)\cr
=&\hbox{${i\over2}$}\h(2\k+\h \G)\s_pg(x)\de'(z-x)-
\hbox{${i\over2}$}\h \beta \s_p g'(x)\de(z-x).\cr
}\eqno(B.5)$$
Therefore  eqs. (B.4) are satisfied only if
$$ \beta=2\k+\h \G.\eqno(B.6)$$
That is, if we define $g(x)$ with $\beta$ given by eqn. (B.6) we find
that eqs. (B.4) are valid.
If we select an alternative value of $\beta$ to that given by the above
equation, the solution to eqn. (B.5) will be non-local  (except in
the abelian case).
Similarly, we can derive eqs. (3.10) regarding the normal ordered
$g^{-1}(x)$ operator.

Finally, we derive eqn. (3.12) for $[T(z),g(x)]+[g(z),T(x)]$. Consider
$$\eqalign{
[t(z),g(x)]=&\e^{ab}[J_a^+(z)J_b^+(z)+J_a^-(z)J_b^-(z)+2J_a^-(z)
J_b^+(z),g(x)]\cr
              =&\e^{ab}\Bigl\{2[J_a^+(z),g(x)]J_b^+(z)+2J_a^-(z)
[J_b^-(z),g(x)]\cr
             &\,\,+  2J_a^-(z)[J_b^+(z),g(x)]+
2[J_a^-(z),g(x)]J_b^+(z)  \cr
 &\,\, +\bigl[J_a^+(z),[J_b^+(z),g(x)]\bigr]-\bigl[J_a^-(z),
[J_b^-(z),g(x)]\bigr]\Bigr\}\cr
    =&-i\h\left[\s^a g(x)J_a^+(x)+\s^a J_a^+(x)g(x)\right]\d(z-x)\cr
     &-{\h^2c\over4}g(x)\left\{[{\d}^+(z-x)]^2-[{\d}^-(z-x)]^2
\right\},\cr}
\eqno(B.7)$$
where   $c$ is defined by eqn. (3.13).
Thus
$$[T(z),g(x)]+[g(z),T(x)]=-{i\h^2c\over{8\pi(2\k+\h \G)}}\Bigl[2g(x)
\de'(z-x)+g'(x)\de(z-x)\Bigr].\eqno(B.8)$$
The corresponding result for $g^{-1}(x)$ (eqn. (3.14)) follows
similarly.

\line{\hfill}

 \noindent {\bf Appendix C: Definition and Form of}
$\sigma_{a}$ {\bf and} $\sigma^{a}$.  \vskip 0.2truecm \noindent
It will be convenient to  define the lower-case base-elements of the centre
$G_c$ of $G_o$ by the equation
$$[\sigma_{a},M_{\pm}]=\pm M_{\pm}^{(a)}  \eqno(C.1)$$
The explicit form of these matrices is then given by
$$\sigma_{a}=I_1+...+I_\alpha-{(n_1+...+n_\alpha)\over N}{\bf
I},\quad(\alpha=a).\eqno(C.2) $$
Here $\bf I$ is the $sl(N)$ identity matrix, and $I_\alpha$ is the
restriction of $\bf I$ to the $sl(n_\alpha)$ block, so that
${\bf I}=\sum_\alpha I_\alpha$.
The dual matrices $\sigma^{a}$ in the centre of $G_o$ are then
$$\sigma^a={I_\alpha\over{n_\alpha}}-{I_{\alpha+1}\over{n_{\alpha+1}}},
\quad(\alpha=a).\eqno(C.3)$$
It is clear from eqn. (C.2) that we have
$$ M_o=\sum_a\s_a.
   \eqno(C.4)$$

\line{\hfill}

\centerline{\bf References}

\line{\hfill}

[1] T.Hollowood and P.Mansfield, Nucl. Phys. {\bf B330} (1990)
720.

[2] P.Bouwknecht and K.Shoutens, Phys. Rep. {\bf C223} 183 (1993).

[3] L.Feher, L.O'Raifeartaigh, P.Ruelle, I.Tsutsui and A.Wipf,
 Phys. Rep. {\bf 222} 1

 $\quad\,$ (1992), and references therein.

[4] A.Bilal and J-L.Gervais, Phys. Lett. {\bf B206} 412 (1988).

[5] J.de Boer and T.Tjin, Commun. Math. Phys. {\bf 160} 183 (1993).

[6] A.Leznov and M.Savilev, Commun. Math. Phys. {\bf 74} 111 (1980).

[7] E.Witten, Commun. Math. Phys. {\bf92} 455 (1984).

[8] P.Goddard and D.Olive, Int. Journal of Modern
Physics {\bf A1} 303 (1986), and

$\quad$ references therein.

[9] L.Faddeev, Commun. Math. Phys. {\bf 132} 131 (1990).

$\quad\,$ J.Balog, L.Dabrowski and L.Feher, Phys. Lett.
{\bf B244} 227 (1990).

[10] V.G.Knizhnik and A.B.Zamolodchikov, Nucl. Phys. {\bf B247} 83
(1984).

[11] T.Curtright and C.Thorn, Phys. Rev. Lett. {\bf 48}, 1309 (1982).

[12] E.D'Hoker and R.Jackiw, Phys. Rev. {\bf D26}, 3517 (1982).

\end